\documentclass[12pt,preprint]{aastex}

\slugcomment{Not to appear in Nonlearned J., 45.}

\shorttitle{Line Diagnostics of AGN Molecualr Torus}
\shortauthors{Yamada et al.}

\begin{document}

\title{HCN to HCO$^{+}$ Millimeter Line Diagnostics of AGN
  Molecular Torus I : \\
  Radiative Transfer Modeling}

\author{Masako YAMADA\altaffilmark{1}}
\affil{ALMA Project Office, % and Division of Theoretical Astronomy,
  National Astronomical Observatory of
  Japan, Mitaka, Osawa, 181-8588, JAPAN}
\email{masako.yamada@nao.ac.jp}

\and
\author{Keiichi Wada and Kohji Tomisaka}

\affil{Division of Theoretical Astronomy, National Astronomical
  Observatory of Japan, Mitaka, Osawa, 181-8588, JAPAN}

%\altaffiltext{1}{present address: ALMA Project Office, National
%  Astronomical Observatory of Japan, Mitaka, Osawa, 181-8588, JAPAN}

%--
\begin{abstract}
We explore millimeter line diagnostics of an obscuring molecular torus
modeled by a hydrodynamic simulation with three-dimensional non-local
thermodynamic equilibrium (nonLTE) radiative transfer calculations.
Based on the results of high-resolution hydrodynamic simulation of the
molecular torus around an active galactic nucleus (AGN), we
calculate intensities of HCN and HCO$^{+}$ rotational lines as two
representative high density tracers.
The three-dimensional radiative transfer calculations shed light on
a complicated excitation state in the inhomogeneous torus, even
though a spatially uniform chemical structure is assumed.
We find that similar transition coefficients for HCN and
HCO$^{+}$ rotational lines lead to a natural concordance of the level
population distributions of these molecules and the line ratio
$R_{\mathrm{HCN}/\mathrm{HCO}+}\lesssim 1$ for the same molecular
abundance value over two orders of magnitude.
Our results suggest that HCN must be much more abundant than HCO$^{+}$
($y_\mathrm{HCN} \gtrsim 10\times y_{\mathrm{HCO}^{+}}$) 
in order to obtain a high ratio ($R_{\mathrm{HCN}/\mathrm{HCO}+}\sim
2$) observed in some of the nearby galaxies.
There is a remarkable dispersion
in the relation between integrated intensity and column
density, indicative of possible shortcomings of HCN$(1-0)$ and
HCO$^{+}(1-0)$ lines as high density tracers.
The internal structures of the inhomogeneous molecular torus down to
subparsec scale in external galaxies will be revealed by the
forthcoming Atacama Large Millimeter/submillimeter Array (ALMA).
The three-dimensional radiative transfer calculations of molecular
lines with high-resolution hydrodynamic simulation prove to be
a powerful tool to provide a physical basis for molecular line
diagnostics of the central regions of external galaxies.
\end{abstract}

%--

\section{Introduction}

Molecular gas at the center of an active galaxy is a subject of
crucial importance for observational and theoretical studies of galaxy
formation.
Various kinds of molecular lines such as $^{12}$CO, HCN,
H$^{12}$CO$^{+}$, CS, and so on have been detected
in the central regions of a number of nearby Seyfert galaxies
\citep{aalto1995, curran2000, usero2004, kohno2005}, and luminous and
ultra luminous infrared galaxies (LIRGs and ULIRGs : \citealt{gao2004b,
  nakanishi2005, papadopoulos2007, gracia-carpio2007}).
Highly sensitive millimeter telescopes have enabled an 
increasing number of observations of high density tracer lines such as
HCN or H$^{12}$CO$^{+}$, aiming at examination of the density
structure of the central regions of external galaxies
\citep{curran2000,gao2004a, gracia-carpio2006}.
Imaging observations of nearby galaxies show complex structures in
molecular gas in the galactic centers (radius $R\lesssim 1$ kpc), such
as compact cores and outer rings (NGC 1068 : \citealt{planesas1991,
  jackson1993, schinnerer2000, usero2004}, M 51 :
\citealt{kohno1996, scoville1998, matsushita2004} : see for reviews of 
recent results, \citealt{kohno2005, garcia-burillo2006, kohno2007}).
The similar profiles of CO and HCN lines, of which optical
thicknesses are suggested to differ, support a small volume filling
factor of high density molecular gas (mist model; \citealt{solomon1987}).
\citet{nguyen1992} concluded from the observation of 15 external
galaxies that molecular gas in the central region is composed of
numerous clumps smaller than the observational beam size of the IRAM
30m telescope ($\Delta\theta_\mathrm{HPBW} = 26^{\prime\prime}$).

Zooming into a smaller scale, the unified model of active galactic
nuclei (AGN) postulates a compact, obscuring dusty molecular torus
around the central engine (e.g., \citealt{antonucci1985}; hereafter we
refer to the molecular gas at the central region $R\lesssim 100$ pc of
an active galaxy as a ``molecular torus''). 
Current observations have failed to resolve the small, dense regions
within a compact molecular torus of radius $\lesssim 100$ pc.
The dynamical and thermal properties of the putative molecular torus,
its structure and origin remain ambiguous.

High resolution hydrodynamic simulations predict a highly
inhomogeneous structure inside the molecular torus, which is
characterized by clumps and filaments on a subpc to pc scale
\citep{wada2001, wada2002, wada2005}.
In such an inhomogeneous gas, the energy level population should also 
become complicated, and interpretation of molecular line
emissions require detailed and careful modeling and calculations. 
For example, numerical simulation of $^{12}$CO rotational lines of
\citet{wada2005} demonstrated a large scatter of ``X-factor'', a
conversion constant between the $^{12}$CO luminosity and molecular
hydrogen column density in an inhomogeneous torus.
Inhomogeneous density and temperature structures are not the unique
factors that generate the intricate excitation conditions. 
There are increasing numbers of arguments concerning the effects of
peculiar chemical evolution by Xray and UV radiation from a central
AGN and accompanying nuclear starburst (\citealt{kohno2001, kohno2005,
  imanishi2006a, imanishi2006, imanishi2004, aalto2007,
  meijerink2007, kohno2007}; see for a
recent review, \citealt{garcia-burillo2006} and references therein).
The observed line intensity is the consequence of the complex coupling
of various factors, such as hydrodynamic and thermal structures, and
chemical abundance distribution.
Furthermore, line intensity of high density tracers, such as HCN, may not
necessarily represent local density \citep{gracia-carpio2006,
  gracia-carpio2007, papadopoulos2007}.

In this article we study the role of high density tracers by the
combination of high resolution hydrodynamic simulation of an
inhomogeneous molecular torus and three-dimensional radiative transfer
calculations without the assumption of local thermodynamic equilibrium
(nonLTE).
We calculate rotational lines of HCN and H$^{12}$CO$^{+}$ (hereafter we
omit the isotope number 12) and
thoroughly examine the excitation status inside the inhomogeneous
molecular torus.
As the first step, we pursue the nonlinear response of nonLTE level
population in the uniform chemical abundance torus. 
The chemical evolution in the torus will be discussed separately in
the subsequent article.

It should be noted that our calculations can present a direct
prediction of the intensity distribution of high density tracer
lines. 
Our radiative transfer results are able to bear direct comparison
with observational data.
They can provide not only valuable suggestions concerning the role of
high density tracers (such as HCN and HCO$^{+}$ lines) in current
observations, but also the theoretical basis for line diagnostics
of the thermal and chemical structure of a molecular torus.
A compact and inhomogeneous molecular torus at the galactic
center will be an interesting target for future observational
instruments, such as the Atacama Large Millimeter/submillimeter Array
(ALMA).
Three-dimensional simulations of a radiation field with high
resolution hydrodynamic simulations will be an important tool for
deriving the astrophysical properties from the
current and forthcoming high quality observational data.

The organization of this article is as follows: 
In \S2 we describe the methods of hydrodynamical and
radiative transfer calculations.
We display our results of the excitation conditions, the intensity and
the line ratio distributions, and the relation with the torus
structure in \S3.
We briefly discuss the possible effect of chemical evolution on
the line ratio of HCN and HCO$^{+}$, and show implications 
for future observations with ALMA in \S4, and summarize our article in
\S5.

\section{Models \& Equations}
\subsection{Hydrodynamic Model of an AGN Molecular Torus}

We first performed hydrodynamic modeling and simulations of the
interstellar medium (ISM) around a supermassive black hole following
\citet{wada2005}.
An ordinary set of hydrodynamic equations is solved with AUSM
\citep[advection upstream splitting method: ][]{liou1993}.
We calculated a $64^2 \times 32$ pc$^3$ region with $256^2\times 128$
uniform Cartesian grid points.
We employed ``model A'' in \citet{wada2005}, in which the molecular
torus evolves in a steady gravitational potential composed of a
supermassive black hole (SMBH) of $M_\mathrm{BH}= 10^8 M_\sun$ at the
center of the galaxy and the galactic dark halo.
These potential fields are described as $\Phi_\mathrm{ext} =
-(27/4)^{1/2}v_c^2 /(r^2+a^2)^{1/2}$ for the former and
$\Phi_\mathrm{SMBH} = -GM_\mathrm{BH}/(r^2+ b^2)^{1/2}$ for the
latter, respectively.
We took the values of the maximum circular
velocity $v_c$ = 100 km s$^{-1}$, the core radius $a= 10$ pc as for the
gravitational potential of galactic dark halo $\Phi_\mathrm{ext}$, and
the core radius $b$ = 1 pc for the SMBH potential
$\Phi_\mathrm{SMBH}$, respectively.
The thermal evolution of the torus was calculated with an empirical
formula of the cooling function of solar metallicity ($Z =
1Z_\sun$), photoelectric heating by strong UV background radiation
($G_0=10$ in Habing unit), and heating by supernova explosions within
the torus.
The parameters that determine the cooling and heating rates,  $Z$ and
$G_0$, are not expected to significantly affect the thermal structure
of the turbulent multi-phase medium in a torus, since they are
determined not only by cooling/heating rates but also by dynamical
processes, and energy feedback from star formation \citep{wada2001,
  wada2002b}.
The torus model assumed a nuclear starburst, which
induces an average supernova explosion rate of $\sim$ 0.36 yr$^{-1}$.
In this model, supernova explosions take place randomly in the
disk plane.
The dynamical evolution of the SNe blast waves in the differential
rotation in the inhomogeneous torus was explicitly calculated in the
simulation, as well as thermal evolution due to radiative cooling.

The structures of density, temperature, and the geometry of the torus
are determined by the balance of the external gravitational forces
(originating in $\Phi_\mathrm{ext}$ and $\Phi_\mathrm{SMBH}$),
self-gravity of the gas, turbulent energy dissipation due to
radiative cooling, and feedback from supernova explosions.
\citet{wada2005} showed that the molecular torus model is globally
quasi-steady over a timescale of $\sim 10^8$ yr, although local
gravitational instability develops a significantly inhomogeneous and 
clumpy internal structure.
In our radiative transfer calculations, we use density, temperature
and velocity structures taken from a snapshot of
the hydrodynamic simulation results.
Strong heating by supernova explosions generates a hot ($T\lesssim
10^6$ K) and tenuous ($n \ll$ 1 cm$^{-3}$) atmosphere above the dense torus
\citep{wada2001, wada2002}.
In the hot atmosphere, neither HCN nor HCO$^{+}$ molecules are expected.
Therefore we remove the hot atmosphere from the hydrodynamic
simulation results as an input to the radiative transfer calculations.
The threshold temperature for removing the hot atmosphere is assumed 
to be $T_\mathrm{th}$ = 1000 K, after comparing with preliminary
radiative transfer calculations with lower $T_\mathrm{th}$ = 400 K.
We found little difference in these results, and then confidently
chose $T_\mathrm{th}=1000$ K.
Thus the temperature range of the input data is 20 K $\le T\le$ 1000
K, and the maximum density reaches as high as $n_{\mathrm{H}_2}
\lesssim 2\times 10^6$ cm$^{-3}$.

%----
\subsection{Radiative Transfer Calculations}

We then calculate HCN and HCO$^{+}$ molecular line intensities using the
results of hydrodynamic simulations. 
The input hydrodynamic data is appropriately smoothed from the
original high resolution results ($256^2\times 128$) to $64^2\times
32$ grid data.
The three-dimensional nonLTE radiative transfer scheme is adopted by
\citet{wada2005} for CO rotational lines.
We calculate nonLTE population distribution simultaneously with
ray-tracing along randomly sampled rays toward the outer boundary
from each cell \citep[for details, see][]{hogerheijde2000}.

The nonLTE level population $n_J$ of the energy level $J$ (where $J$ is
the rotational quantum number) is
calculated by solving the statistical equilibrium rate equation, 
  \begin{eqnarray}
    n_J\sum_{J^{\prime}\ne J}R_{J J^{\prime}} &=& 
       \sum_{J^{\prime}\ne J}(n_{J^{\prime}}R_{J^{\prime}J}),
       \label{eq:pop} \\
    R_{JJ^{\prime}} &=& \cases{ 
                A_{JJ^{\prime}}+B_{JJ^{\prime}}\bar{J}
		   +C_{JJ^{\prime}} & $J > J^{\prime}$,  \cr
                B_{JJ^{\prime}}\bar{J} + C_{JJ^{\prime}} & $J < J^{\prime}$,
               \cr }  \label{eq:pop2}
  \end{eqnarray}
where $A_{JJ^{\prime}}$, $B_{JJ^{\prime}}$ represent the Einstein's
coefficients from the energy level $J$ to $J^{\prime}$, and
$C_{JJ^{\prime}}$ is the collisional transition rate per unit time,
respectively.
In equation (\ref{eq:pop}), the left hand side is the outgoing rate
from level $J$ under consideration, and the right hand side is the
incoming transition rate to level $J$.
The average intensity $\bar{J}$ is calculated from the specific
intensity $I_\nu$,
  \begin{equation}
    \bar{J} \equiv
    \frac{1}{4\pi}\int\int_0^{\infty}I_{\nu}\phi(\nu)d\nu 
     d\Omega,
    \label{eq:jbar}
  \end{equation} 
where $\phi(\nu)$ is the normalized absorption coefficient profile
(see below).
The collisional transition rate $C_{JJ^{\prime}}$ is described as
  \begin{equation}
    C_{JJ^{\prime}} = \sum_X \gamma_{JJ^{\prime}} n_X, \nonumber
  \end{equation}
where $n_X$ represents the number density of collision
partner, and $\gamma_{JJ^{\prime}}$ is the corresponding collisional
excitation (or de-excitation) coefficient.
In this paper we replace $n_X$ by the number density of molecular
hydrogen $n_{\mathrm{H}_2}$.

We calculate the specific intensities $I_{\nu}$ by solving the
equation of radiative transfer, 
  \begin{equation}
    \frac{dI_{\nu}}{d\tau} = -I_{\nu}+S_{\nu},  \label{eq:rad}
  \end{equation}
where $S_\nu$ denotes the source function.
Radiatively induced transition rates are calculated from $\bar{J}$,
which is obtained by averaging the specific intensities over the
sampling rays coming to each grid cell based on the Monte-Carlo
method.
Rate equations (\ref{eq:pop}), (\ref{eq:pop2}), and transfer
equation (\ref{eq:rad}) are iteratively solved until both the level
population and radiation field converge self-consistently. 
The convergence speed is improved by a version of Accelerated
Lambda Iteration (ALI, \citealp{hogerheijde2000}) to be optimizedly tuned
for our scalar-parallel computer.
At the outer boundary we impose the radiation field to be identical to
the Cosmic Microwave Background radiation (CMB).

As for transition coefficients ($A_{JJ^{\prime}}$, $B_{JJ^{\prime}}$,
and $\gamma_{JJ^{\prime}}$), we adopt the database of Leiden
University \citep{schoier2005}.
\footnote{LAMBDA : http://www.strw.leidenuniv.nl/\texttt{\~}moldata/}
In order to obtain accurate level population distributions, we
solve the rate equation for $0\le J\le J_\mathrm{max}$ with the
maximum energy level $J_\mathrm{max} =$ 10 for both HCN and
HCO$^{+}$ lines.
The number of sampling rays for $\bar{J}$ ranges from $\sim 300$ to
$\sim 900$ for each grid point to achieve a convergence level $|\Delta
n_J/n_J|_i \lesssim 10^{-6}$ for all $J$ levels ($\Delta n_J|_i
= n_J^i-n_J^{i-1}$ is defined as the maximum of the difference between
$n_J$ of the $i$-th and $(i-1)$-th iterations on the same grid).
For the low energy level $J$ ($J\lesssim$ 7), the degree of
convergence progressively improves to $\Delta n_J/n_J \lesssim 10^{-10}$.

The model torus has a supersonic turbulent velocity field of a large 
dispersion $\Delta v\lesssim$ 50 km s$^{-1}$, in comparison with
the thermal velocity 0.2 km s$^{-1} \le c_s \le$ 6.4 km s$^{-1}$.
This turbulent velocity influences the degree of ``overlapping'' of
the line which accounts for the optical thickness.
We assume microturbulence in the absorption coefficient profile
$\phi(\nu)$, 
  \begin{eqnarray}
    \phi(\nu)d\nu &=& \frac{1}{\Delta \nu \sqrt{\pi}}
      \exp \left[ -\frac{(\nu -
      \nu_0)^2}{(\Delta \nu)^2}\right]d\nu,
      \label{eq:lprofile}  \\
    \int^{+\infty}_0 \phi(\nu)d\nu &=& 1,
  \end{eqnarray}
where $\nu_0$ is the line center frequency measured in the rest frame
of each fluid element, and $\Delta \nu \equiv (\nu_0/c)\cdot
\sigma_\mathrm{turb}$ is the effective Doppler width described by
$\sigma_\mathrm{turb}$, the velocity of microturbulence.
The velocity structure of the torus is reflected via 
the profile $\phi(\nu)$ in equation (\ref{eq:lprofile}).
In order to determine a reasonable value for $\sigma_\mathrm{turb}$, we
survey the parameter space ranging between 1 km s$^{-1}$ $\le
\sigma_\mathrm{turb} \le$ 50 km s$^{-1}$. 
After confirming the convergence stability of the solution over a wide
range of parameters 10 km s$^{-1} \lesssim \sigma_\mathrm{turb}
\lesssim$ 50 km s$^{-1}$, we choose the value of  $\sigma_\mathrm{turb}
=$ 20 km s$^{-1}$ in the following calculations.

NonLTE population calculation starts with solution of the rate
equation (\ref{eq:pop2}) in the optically thin limit only with 
background radiation $\bar{J} = B_{\nu} (T_\mathrm{CMB})$ (where
$B_\nu(T)$ denotes the Planck function) throughout the torus.
We examine the effects of the initial guess of the population
distribution on the final solution by starting with an alternative
initial condition.
As for an opposite extreme alternative, we select the Boltzmann
level distribution with the local kinetic temperature of each grid for a
starting point.
We confirm that these two initial conditions result in the same
solution within errors of the order $|\Delta n_J/n_J| \lesssim
10^{-5}$.
Convergence speed of the radiative transfer calculation is about 2 to
3 times faster for the optically thin initial condition than the LTE
for both HCN and HCO$^{+}$ lines.
We conclude that both initial conditions give correct solutions,
and then we adopt the optically thin initial condition in our
radiation calculations because of its faster convergence to the final
solution.

In order to focus on the effects of inhomogeneous torus structure on
the excitation conditions and emergent line intensities, we assume
spatially uniform abundances of emitting molecules (HCN
and HCO$^{+}$).
The fractional abundance $y\equiv n_\mathrm{mol}/n_{\mathrm{H}_2}$ is
assumed to be $10^{-11}\le y \le 10^{-7}$.
For the same reason, we ignore the radiative pumping by
continuum emission from AGN and/or nuclear starbursts.
The non-uniform abundance distribution due to AGN and nuclear
starburst will be separately discussed in paper II (Yamada et al., in
prep).

%----
\section{Results}

%----
\subsection{Intensity Distribution} \label{sect:intdist}

\begin{center}
  \begin{figure}
    \includegraphics[width=17cm,clip]{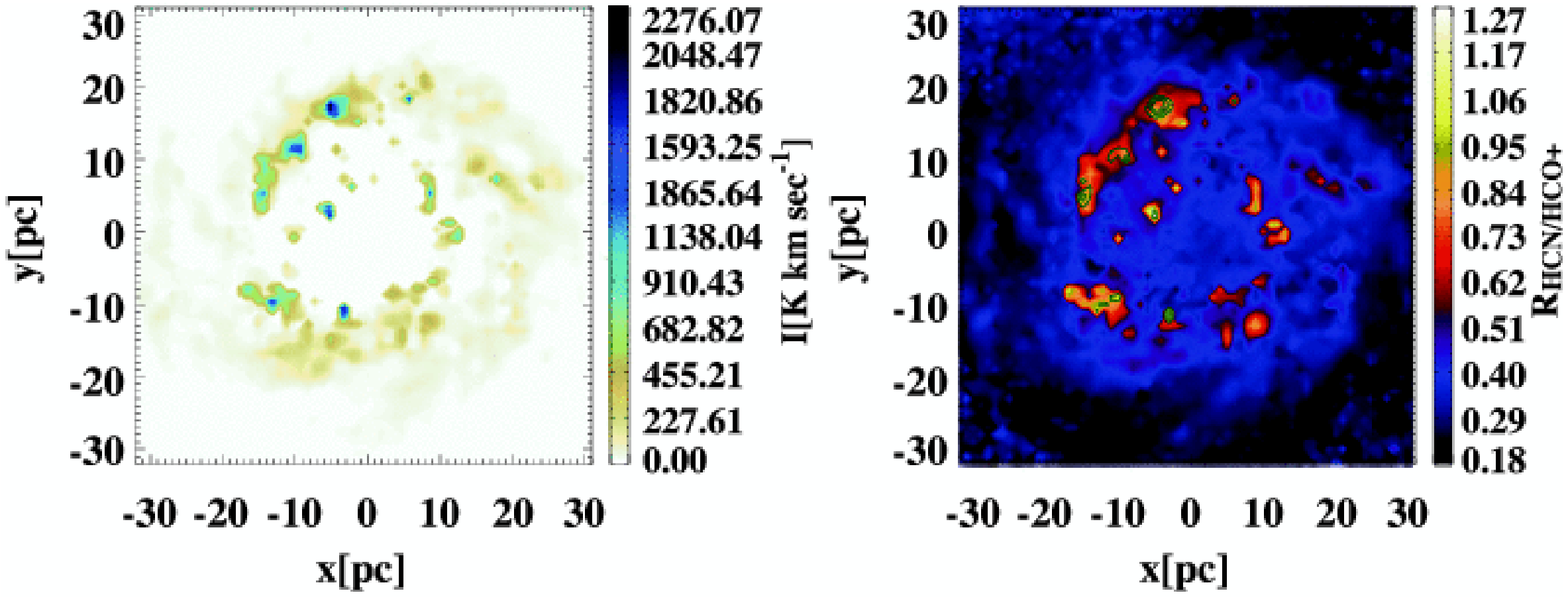}
    \caption{Left panel: integrated flux distribution of HCN$(1-0)$
    line with $y_\mathrm{HCN} = 2\times 10^{-9}$ for face-on data.
    Right panel: the distribution of the line ratio
    $R_{\mathrm{HCN/HCO}^{+}}$ with the same abundance value of $y$.
    In the right panel, the maximum number density of hydrogen
    molecules on all the lines of sight are overplotted with green
    contours.
    }
    \label{fig:intmap}
  \end{figure}
\end{center}

The left panel of Figure \ref{fig:intmap} displays the integrated
intensity distribution of HCN$(1-0)$ line of a face-on torus.
In this panel we adopt the fiducial value of $y_\mathrm{HCN} = 2\times
10^{-9}$ for the molecular abundance. 
The panel exhibits a significantly inhomogeneous intensity
distribution which reflects the highly inhomogeneous structure inside
the torus (see Figs. 1 and 2 of \citealp{wada2005} for the
density structure).
Since level population is independent of the viewing angle,
hereafter we use the face-on data in the following analysis unless
otherwise stated.

The right panel of Figure \ref{fig:intmap} displays the line
ratio distribution of HCN$(1-0)$ and HCO$^{+}(1-0)$, 
$R_{\mathrm{HCN(1-0)/HCO}^{+}(1-0)} \equiv (\int T_b(\mathrm{HCN})dv)
/ (\int T_b(\mathrm{HCO}^{+}) dv) = R_{\mathrm{HCN/HCO}^{+}}$ ($T_b$
is the brightness temperature) for the same molecular abundance
$y_\mathrm{HCN} = y_{\mathrm{HCO}^{+}} = 2\times 10^{-9}$.
The panel shows that in spite of the significant inhomogeneity of the
intensity distributions, the line ratio is restricted in a narrow 
range from $\sim 0.2$ to $\sim 1.2$ (the probability distribution function of
$R_{\mathrm{HCN/HCO}^{+}}$ has a sharp peak around the median $\sim
0.6$). 
The similarity of HCN$(1-0)$ and HCO$^{+}(1-0)$ intensities arises
from the almost identical values of rotational constant $B$ and
permanent electric dipole moment $\mu_e$ for these molecules,
which determine the energy levels and Einstein's $A$ and $B$
coefficients for pure rotational transitions (see Table
\ref{tbl:coef}).
The coefficients in Table \ref{tbl:coef} are calculated by the
equations 
\begin{eqnarray}
  E_J &=& BhJ(J+1), \\
  B_{J,J-1} &=& \frac{32}{3} \pi^4 \mu_e^2 \frac{1}{h^2c}
         \frac{J}{2J+1},   \label{eq:einsB} \\
  A_{J,J-1} &=& 16 \frac{h}{c^2} B^3J^3 B_{J, J-1},  \label{eq:einsA}
\end{eqnarray}
where $E_J$ is the energy measured from the ground state, $h$ is the
Planck constant, and $c$ is the velocity of light.
Besides these coefficients, the critical density for LTE
distribution, $n_\mathrm{crit}\equiv
A_{J,J^{\prime}}/\gamma_{J,J^{\prime}}$ is different between HCN and
HCO$^{+}$ molecules.
However, the amount of gas mass of densities larger than the
critical density of HCN$(1-0)$ ($n_\mathrm{crit}\sim 10^6$
cm$^{-3}$) is $\sim 3.6\times 10^5 M_{\sun}$, and that larger than
critical density of HCO$^{+}(1-0)$
($n_\mathrm{crit}\sim 10^5$ cm$^{-3}$) is $\sim 1.3\times 10^6
M_{\sun}$, which agree within a factor of order unity.
In the optically thin limit, line ratio straightforwardly traces the
mass ratio, which becomes about unity in our simulation. In our
calculations average optical thickness over the entire field of view
is at most $\langle\tau_0\rangle = 2$ (see \S3.3 below), and thus
dependence of line ratio on critical densities is weak.

\begin{center}
  \begin{figure}
    \includegraphics[width=10cm]{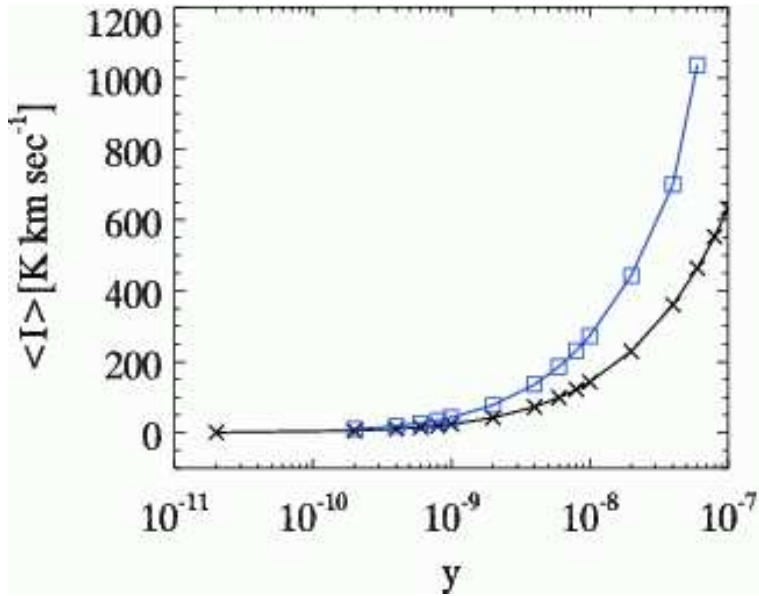}
    \caption{Spatially averaged integrated intensity $\langle
    I\rangle$ over the entire field of view as a function of molecular
    abundance $y$.
    A line with squares displays $\langle I\rangle$
    of HCO$^{+}(1-0)$ emission, and that with crosses presents
    $\langle I\rangle$ of HCN$(1-0)$ emission, respectively.}
    \label{fig:abund_u777}
  \end{figure}
\end{center}

The overall tendency $R_{\mathrm{HCN/HCO}^{+}} \simeq \mathcal{O}(1)$
does not depend on the molecular abundance $y$.
In Figure \ref{fig:abund_u777} we present the integrated intensities
averaged over the field of view (64$\times$ 64 pc$^2$) as a function
of $y$ for both HCN$(1-0)$ and HCO$^{+}(1-0)$ lines.
Figure \ref{fig:abund_u777} shows that
$\langle I\rangle_{\mathrm{HCN}(1-0)}(y) \lesssim
\langle I\rangle_{\mathrm{HCO}^{+}(1-0)}(y)$ for over 2 orders of
magnitude of $y$. 
This inequality
$\langle I\rangle_{\mathrm{HCN}(1-0)}(y)/\langle
I\rangle_{\mathrm{HCO}^{+}(1-0)}(y) <
1$ implies that in order to obtain a high ratio
$R_{\mathrm{HCN}/\mathrm{HCO}^{+}} \sim 2$ observed in some 
nearby galaxies \citep[e.g., NGC 1068, ][]{usero2004}, 
HCN molecules should be much more abundant than HCO$^{+}$ molecules
($y_\mathrm{HCN} \gg y_{\mathrm{HCO}^{+}}$).
For example, if we take $y_{\mathrm{HCO}^{+}} = 2\times 10^{-8}$,
Figure \ref{fig:abund_u777} means that $y_\mathrm{HCN}$ should be as
large as $\gtrsim 10^{-7}$ (or $y_\mathrm{HCN}\gtrsim 10\times
y_{\mathrm{HCO}^{+}}$) for a high ratio
$R_{\mathrm{HCN}/\mathrm{HCO}^{+}} \simeq 2$.
\begin{table}
\begin{center}
  \begin{tabular}{|c|c|c|c|c|}\hline
    mol. & $ 2B  $[GHz] & $\mu_e$ & $A_{10}$ [Hz] & $B_{10}$  \\ \hline
    HCN  & 88.63 & 2.99 $\times 10^{-18}$ 
         & 2.40 $\times 10^{-5}$ & $2.35\times 10^{9}$  \\
    HCO$^+$ & 89.19 & 3.93 $\times 10^{-18}$ 
         &  4.25 $\times 10^{-5}$ & $4.25\times 10^{9}$  \\
    \hline
  \end{tabular}
  \caption{Fundamental parameters of HCN and HCO$^{+}$ pure rotational
  transitions.
  The value $2B$ is twice the rotation constant and is identical to
  $\nu_{10}$.
  Electric dipole moment $\mu_e$ is measured in cgs units, $A_{10}$
  measures the number of spontaneous transitions per unit
  time, $B_{10}$ is the stimulated radiative transition coefficient
  per unit time and per unit intensity, respectively.}
  \label{tbl:coef}
\end{center}
\end{table}

%------
\subsection{The Line Ratio $R_{\mathrm{HCN}/\mathrm{HCO}^{+}}$ in the
  Bright Regions} \label{sect:intweight}

The left panel of Figure \ref{fig:intmap} shows that the torus has many
bright spots of several parsec in size in the integrated intensity
distribution.
The line ratio distribution has a smoother structure with
$R_{\mathrm{HCN}/\mathrm{HCO}^{+}}\approx 1$ (the right panel of
Fig. \ref{fig:intmap}).
Unfortunately the observational beam size of present instruments is
too large to resolve such small structures inside a compact torus (for
instance, $\Delta\theta_\mathrm{HPBW} = 2^{\prime\prime} \sim
6^{\prime\prime}$ for NMA and Rainbow interferometers typically
correspond to several hundred pc except for Local Group members).
Observational estimation of line ratio would be practically governed
by the ratio of the bright regions within the unresolved torus.
In the following paragraph, we examine the line ratio
$R_{\mathrm{HCN}/\mathrm{HCO}^{+}}$ in the bright region, which is
expected to be close to the observational estimation in the unresolved
molecular gas.

\begin{center}
  \begin{figure}
    \includegraphics[width=17cm,clip]{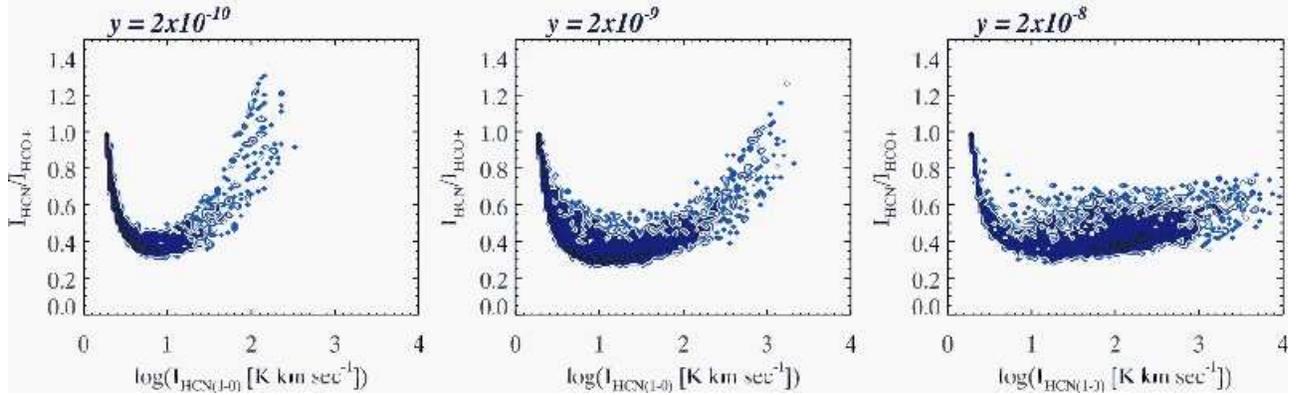}
    \caption{Combined probability distribution function of the
      integrated intensity and the ratio
      $R_{\mathrm{HCN}/\mathrm{HCO}^{+}}$.
      Abscissa denotes the integrated intensity of HCN $(1-0)$ line
      and ordinate denotes the ratio of
      $I_{\mathrm{HCN}(1-0)}/I_{\mathrm{HCO}^{+}(1-0)}$ per pixel,
      respectively.
      Three panels correspond to different abundances $y$, $2\times
      10^{-10}$, $2\times 10^{-9}$, and $2\times 10^{-8}$ from left to
      right.
      For a small value of $y$, the line ratio
      $R_{\mathrm{HCN}/\mathrm{HCO}^+}$ becomes $\simeq 1$ in all
      pixels.
      As $y$ increases, though, the number of pixels of low ratio
      $R_{\mathrm{HCN}/\mathrm{HCO}^+} \le 1$ accordingly
      increases in bright pixels.
    }
    \label{fig:int_ratio}
  \end{figure}
\end{center}

We show two-dimensional probability distribution function of
the integrated intensity $I(x, y)$
and the ratio $R_{\mathrm{HCN}/\mathrm{HCO}^{+}}(x, y) =
I_{\mathrm{HCN}}(x,y)/I_{\mathrm{HCO}^{+}}(x,y)$ in Figure
\ref{fig:int_ratio} (where $I(x, y)$ is the integrated intensity at
the position $(x, y)$ in the field of view).
Three panels in Figure \ref{fig:int_ratio} are for different molecular
abundances, $y=2\times 10^{-10}$, $2\times 10^{-9}$, and $2\times
10^{-8}$, respectively.
Figure \ref{fig:int_ratio} shows the line ratio
$R_{\mathrm{HCN}/\mathrm{HCO}^{+}}\lesssim 1$  for almost all the
pixels irrespective of the integrated intensity $I(x, y)$.
When $y$ is very small ($ = 2\times 10^{-10}$), pixels of high ratio
up to $R_{\mathrm{HCN}/\mathrm{HCO}^{+}}\lesssim 1.4$ remain at
the bright end.
As $y$ approaches a realistic abundance value $y=2\times 10^{-8}$
inferred by observational results (e.g. NGC 1068,
\citealt{usero2004}), however, the line ratio at the bright end
falls below $R_{\mathrm{HCN}/\mathrm{HCO}^{+}}\lesssim 0.8$.
The lower $R_{\mathrm{HCN}/\mathrm{HCO}^{+}}$ towards the
brighter points in the case of $y=2\times 10^{-8}$ may indicate the
possible underestimation of the line ratio
$R_{\mathrm{HCN}/\mathrm{HCO}^{+}}$ in the current observations of
spatially unresolved tori.
This trend in Fig.\ref{fig:int_ratio} might even strengthen the
necessity for abundant HCN over HCO$^{+}$ to obtain a high ratio
($R_{\mathrm{HCN}/\mathrm{HCO}^{+}}\sim 2$).
On the other hand, it might be possible that different spatial
distributions of HCN and HCO$^{+}$ molecules loosen 
this requirement, but this topic is out of this paper and will 
be discussed in a subsequent paper.

%----------
\subsection{The Population Distribution in the Radiative
  Transfer Calculations}

In this subsection we demonstrate the population distribution of
our nonLTE calculation results (the basic characteristics of multi-level
population distribution in statistical equilibrium in the optically
thin limit are summarized in Appendix).
In Figure \ref{fig:fj10_sim}, we plot the ratio $n_1/n_0$ against
density $n_{\mathrm{H}_2}$ for three models with $y=2\times 10^{-10}$,
$2\times 10^{-9}$, and $2\times 10^{-8}$.
In Figure \ref{fig:fj10_sim}, the level population in the optically thin
limit is also shown with lines.
Figure \ref{fig:fj10_sim} exhibits two distinct features of the ratios 
$n_1/n_0$ of radiative transfer calculations compared with 
optically thin cases: 1) larger values of $(n_1/n_0) \gtrsim
(n_1/n_0)_\mathrm{thin}$ on the low density side ($n_{\mathrm{H}_2}
\lesssim 10^4$ cm$^{-3}$), and 2) smaller peak value of
$(n_1/n_0)_\mathrm{peak} < (n_1/n_0)_\mathrm{peak, thin}$ at
$n_{\mathrm{H}_2}\sim n_\mathrm{crit}$. 
The departure from optically thin distribution escalates with the
molecular abundance $y$.
These features on the diagram can be elucidated by 
the increase in the radiation-induced transition rate ($\propto
\bar{J}$) by the propagation of line emission in the torus.

The effect of frequent radiation-induced transitions appears on the
diagrams as 1) the broad dispersion in the $n_1/n_0$
distribution on the low density side ($10^2$ cm$^{-3} \lesssim
n_{\mathrm{H}_2} \lesssim 10^4$ cm$^{-3}$), and 2) the decrease in the
local peak value of $n_1/n_0$ at the density $n\sim n_\mathrm{crit}$.
In the low density regime ($n\ll n_\mathrm{crit}$), collisional
transitions are negligible and the rate equation becomes 
  \begin{equation}
    n_0B_{01} \bar{J} = n_1A_{10} +
    n_1B_{10}\bar{J}, \label{eq:a_thin}
  \end{equation}
and the ratio $n_1/n_0$ is written as
  \begin{equation}
    \frac{n_1}{n_0} =
    \frac{g_1/g_0B_{10}\bar{J}}{A_{10}+B_{10}\bar{J}}.
       \label{eq:thinlim}
  \end{equation}
Equation (\ref{eq:thinlim}) means that $n_1/n_0$ 
increases with the average intensity $\bar{J}$, and can become as
large as $n_1/n_0 \le g_1/g_0 = 3$.
Therefore $n_1/n_0$ can take a value ranging from 0.67
(when $\bar{J}$ is equal to the background radiation,
$J_\mathrm{CMB}$; see Eq. [\ref{eq:thinlim10}]) to 3 according to
$\bar{J}$.
In the denser regime, the decrement of the local peak value of
$n_1/n_0$ at $n_{\mathrm{H}_2}\sim n_\mathrm{crit}$ is also
explained by the similar argument in Appendix \ref{sect:thinpop} as
follows: 
A large radiative excitation rate ($\propto \bar{J}_{10}\equiv
1/(4\pi)\int\int I_{\nu}\phi(\nu_{10})d\nu d\Omega$) increases the
ratio $n_1/n_0$, and then the excitation rate from the
level $J=1$ (not from the ground level $J=0$) increases as well.
The increase in the excitation rate from level $J=1$ obstructs the
downward cascade $J=2\rightarrow 1$, and thus leads to the reduction
of particle accumulation at $J=1$ (see Appendix \ref{sect:thinpop} for
details).

Increase in molecular abundance $y$ raises the local emission rate
($j_\nu = \alpha_\nu S_\nu$ in Eq. [\ref{eq:rad}] increases as $\propto
n_\mathrm{mol}n_{\mathrm{H}_2}\propto
y_\mathrm{mol}n_{\mathrm{H}_2}^2$), and then the average $\bar{J}$
increases with $y$ as well (Eq.[\ref{eq:jbar}] and [\ref{eq:rad}]).
Hence the radiation-induced transition rate due to $\bar{J}$ increases
with $y$.
The increase in $\bar{J}_{10}$ reduces the peak value of $n_1/n_0$ at
$n_{\mathrm{H}_2}\sim n_\mathrm{crit}$.
The dispersion of $n_1/n_0$ in the low density regime and the degree of
reduction in the peak value of $n_1/n_0$ at $n_{\mathrm{H}_2}\sim
n_\mathrm{crit}$ become larger for a larger $y$ in Figure
\ref{fig:fj10_sim}.
This behavior also supports the importance of radiation induced
transitions ($\propto\bar{J}$) in determining the excitation
conditions in the inhomogeneous torus.

In addition to the increased radiation-induced transitions, the local
photon trapping can also affect the distribution of $n_1/n_0$ for a
large $y$ case.
The effect of the local photon trapping can be described in terms of
the escape probability $\beta$ \citep{goldreich74, peraiah2002}.
The local photon trapping reduces the effective critical density as
$n_\mathrm{crit}(\mathrm{eff.})\approx \beta A_{10}/\gamma_{10}$.
%\approx \tau_0^{-1}A_{10}/\gamma_{10}$.
%
The escape probability $\beta$ is described as
$\beta=(1-e^{-3\tau_0})/3\tau_0$ for a gas slab and
$\beta=(1-e^{-\tau_0})/\tau_0$ for a spherical gas
\citep[e.g,][]{peraiah2002}. 
Though our model torus is neither a slab nor a sphere, both formula
provide limitting values of $\beta\approx 1$ for small $\tau_0$
($\tau_0\lesssim 1$) and $\beta\approx \tau_0^{-1}$ for a large
$\tau_0$ ($\tau_0\gg 1$). 
The average optical thickness in the field of view is estimated to be 
$\langle \tau_0\rangle=$ 2.00 for the face-on data at the line
center of HCN in our simulations for a large $y = 2\times 10^{-8}$
result. 
The average optical thickness of order of unity in our calculations
results in the similar value of $\beta$, so that only a negligible
effect due to $\beta$ is expected on the diagrams of Figure
\ref{fig:fj10_sim}.

%-----
\subsection{The Effects of Inhomogeneous Structure on Optical
  Thickness and Intensity Distributions}

In the previous subsections we have examined the statistical
characteristics of the excitation conditions in the torus.
In order to examine the connection between the resultant emission and
the input torus properties, in Figure
\ref{fig:tau0} we present the integrated intensity
as a function of column density $N_{\mathrm{H}_2}$.
Figure \ref{fig:tau0} shows that while the integrated intensity $I$ is
tightly correlated with $N_{\mathrm{H}_2}$ in the low
$N_{\mathrm{H}_2}$ regime ($N_{\mathrm{H}_2}\lesssim 10^{23}$
cm$^{-2}$), the dispersion of $I$ appears at the larger
$N_{\mathrm{H}_2}$ regime ($N_{\mathrm{H}_2}\gtrsim 10^{23}$ cm$^{-2}$).
The origin of this dispersion is further investigated in the relations 
of integrated intensity, optical thickness, and column density
(Fig. \ref{fig:tau_nh}).
Figure \ref{fig:tau_nh} $(a)$ shows a dispersion in the
optical thickness $\tau_0 = \int \alpha_\nu ds$ at large column
density ($N_{\mathrm{H}_2}\gtrsim 10^{23}$ cm$^{-2}$).
This dispersion is obviously produced by the diversity of the
absorption coefficients $\alpha_\nu(n_{\mathrm{H}_2},T)$ due to the
nonLTE level population in the inhomogeneous torus.
The drastic increase in integrated intensity appearing at $\tau_0\le
0$ (Fig. \ref{fig:tau_nh} $[b]$) owes the stimulated emission by
the population inversion (Appendix \S\ref{sect:popinv}).
\footnote{In our simulation, the gain factor $e^{-\tau_0}$ due to
  population 
inversion is at most a several times of 10 (see
Fig. \ref{fig:tau_nh}$[b]$). 
This gain factor might be underestimated by the adopted 
microturbulence,  $\sigma_\mathrm{turb}= 20$ km s$^{-1}$ in counting
for $\tau_0$ in a grid (Eq. \ref{eq:lprofile}). Higher resolution
simulation might find even stronger masar spots amplified by the
population inversion and velocity coherence, though this kind of
calculation is computationally too hard a task at present.}
In the two panels of Figure \ref{fig:tau_nh}, crosses indicate the values
of the average integrated intensity ($\langle I\rangle = 42.0$ K km
sec$^{-1}$), the average optical thickness ($\langle \tau_0\rangle =
0.37$), and the average column density ($\langle
N_{\mathrm{H}_2}\rangle = 9.27\times 10^{22}$ cm$^{-2}$) in the
face-on field of view.
Our three-dimensional nonLTE radiative transfer calculations reveal
the large dispersions around the average values of $I$, $\tau_0$, and
$N_{\mathrm{H}_2}$.
These results imply that high angular resolution observations of ALMA
are of crucial importance to study the structure of the torus.

The spatial correlations of optical thickness and integrated
intensity also show a wide variety in our results.
In Figure \ref{fig:tau_o1}, we plot the spatial distributions of the
integrated intensity of HCN$(1-0)$ and the optical thickness $\tau_0$.
The disaccordance of the integrated intensity distribution (color map)
and the optically thick region ($\tau_0 \ge 1$) is obvious in this
panel.
This disaccordance is reasonably well explained by the different
dependence of $\tau_0$ and $I$ on density and the line-of-sight
structure of the molecular torus as follows.

In three panels of Figure \ref{fig:tau_o}, the distributions of
$\Delta\tau_0$ (the optical thickness per single grid) and the 
density along three lines of sight are displayed.
The locations of three lines of sight are indicated in
Figure \ref{fig:tau_o1}.
The three lines of sight represent different circumstances in the
inhomogeneous torus.
In panel $(a)$ emission from a clump of $n_{\mathrm{H}_2} \lesssim
10^3$ cm$^{-3}$ is visible through a tenuous and optically thin
ambient medium, in panel $(b)$ a tenuous atmosphere encompasses a
large scale height in the outer torus, and in panel $(c)$ the
stimulated emission due to the population inversion inside a high
density region around $z=0$ plane dominates the intensity.
Since the local emission rate $j_\nu= \alpha_\nu S_\nu$
($\propto n_\mathrm{HCN}n_{\mathrm{H}_2}\propto n_{\mathrm{H}_2}^2$)
and the optical thickness $\tau_0$ ($\propto N_{\mathrm{H}_2} \propto
n_{\mathrm{H}_2}$) are differently dependent on density, the
integrated intensity can be strong even if $\tau_0$ is not large in an
inhomogeneous torus (Fig.\ref{fig:tau_o} $a$ and $c$).
When population inversion ($\alpha_\nu <0$) occurs (in panel $c$),
the integrated intensity can be much larger than that simply
estimated from the column density (Fig.\ref{fig:tau0}).
This is because while the intensity rises rapidly in proportion
to $e^{|\Delta\tau_0|}$, the optical thickness $\tau_0 = \int \alpha_0
ds$ only weakly increases due to the negative $\alpha_0$ in the
integrand.

Our results show that even if the chemical abundance distribution is
uniform, the distribution of the intensity and the optical thickness
can significantly differ.
Hence it would be possible that the line ratio reflects neither 
the ratio of the corresponding amount of the gas of density higher
than critical density, nor the ratio of the area of the $\tau_0=1$
surface.
Conventional arguments about the amount of dense gas or the $\tau_0=1$
surface area are applicable only if the optical
thickness is assured to be quite small ($\tau_0\ll 1$) or large
($\tau_0\gg 1$).
However, if the ISM is subthermally populated and/or has  
$\tau_0\sim 1$ (which would be the case in the molecular torus), one
should take account of the large dispersion of $\alpha_\nu$,
especially of population inversion.

%%%%%%%%%%%%%%%
\section{Discussion}

We performed three-dimensional radiative transfer calculations with 
the assumption of spatially uniform chemical abundance,
and closely examined how the inhomogeneous torus structure affects the
excitation conditions and the line ratio $R_{\mathrm{HCN/HCO}^{+}}$.
Our results in \S3 are briefly summarized as follows:
1) the expected line ratio of the same molecular abundance $y$ is
$R_{\mathrm{HCN/HCO}^{+}}\lesssim 1$ over a wide range of $y$, and 
2) the intensity can be strongly affected by the stimulated emission by
the population inversion.
The arguments in \S\ref{sect:intweight} and Figure \ref{fig:int_ratio}
indicate the smaller line ratio $R_{\mathrm{HCN/HCO}^{+}}\lesssim 1$
in bright regions, and thus the observational estimation of line ratio
of the unresolved torus could not be the ``true'' average ratio
$R_{\mathrm{HCN/HCO}^{+}}$ including the faint regions.
Even if we consider the inhomogeneous thermal structure in the torus, 
a ratio of $R_{\mathrm{HCN/HCO}^{+}}> 1$ is quite
difficult to achieve with the assumption of spatially uniform chemical
abundance (\S\ref{sect:intweight}).

%--
\subsection{HCN and HCO$^{+}$ Line Ratio as a Probe of Chemistry of
  Molecular Gas}

Molecular line intensities could reflect the chemical evolution of
molecular gas under the irradiation of Xray and UV emissions from the
galactic center.
This kind of study began with the pioneering work of \citet{kohno2005}
on the ``HCN diagram'' of nearby Seyfert and starburst galaxies.
His ``HCN diagram'' shows a trend of stronger HCN line emission in
nearby galaxies with little signature of starbursts
($R_{\mathrm{HCN/HCO}^{+}}\gtrsim 2.0$ and $R_\mathrm{HCN/CO}\gtrsim
0.4$), compared with those with starburst regions \citep[see
  also][]{imanishi2006a,imanishi2006,gracia-carpio2006}.
In addition to local Seyfert galaxies, a number of other types of
galaxies show high ratios $R_{\mathrm{HCN/HCO}^{+}}> 1.0$ (for
example, NGC 4418; \citealt{imanishi2004}, UGC 5101 and Mrk 273;
\citealt{imanishi2006}; IC 342, Maffei 2, NGC 2903,
\citealt{nguyen1992} ; NGC 1097, NGC 1068, NGC 5194, \citealt{kohno2005}). 
In this subsection we briefly discuss our radiative transfer results
and chemistry.

Our results show the ratio $R_{\mathrm{HCN/HCO}^{+}} \lesssim 1$ for
the same molecular abundance $y$ over 2 orders of magnitude (the
right panel of Fig. \ref{fig:intmap} and Fig. \ref{fig:int_ratio}).
The arguments in \S\ref{sect:thinpop} derive the reason for  
the convergence of the ratio $n_1/n_0$ of HCN and HCO$^{+}$ in both  
the extremities of the low and high densities in the optically thin
limit (Fig. \ref{fig:f10_den2}).
Since the intensity $I_{10}$ increases as $yn_1$, the
intensity ratio $R_{\mathrm{HCN/HCO}^{+}}$ roughly agrees with the
fractional level population ratio $n_1/n_0$ of HCN and HCO$^{+}$.
Hence for the same value of $y$, line ratio $R_{\mathrm{HCN/HCO}^{+}}$
becomes order of unity (Fig. \ref{fig:int_ratio}).
\footnote{The brightness
  temperature $T_b$ is proportional to $\nu^2$, and therefore
  $R_{\mathrm{HCN/HCO}^{+}}$ is accordingly reduced for the same value
  of $n_1/n_0$.}
This result means that, in order to obtain a high ratio
  $R_{\mathrm{HCN/HCO}^{+}}> 1$,  HCN should be
  much more abundant than HCO$^{+}$ ($y_\mathrm{HCN} \gg
  y_{\mathrm{HCO}^{+}}$). 
Recently \citet{gracia-carpio2007} found that the high line ratio
  requires about 10 times larger HCN fractional abundance than
  HCO$^{+}$ from their Large Velocity Gradient (LVG) analysis of
  $J=3-2$ and $J=1-0$ transitions of LIRGs and ULIRGs.
Our results are approximately consistent with theirs.

Several groups discussed the chemical evolution scenario for a high
ratio $R_{\mathrm{HCN/HCO}^{+}}$.
\citet{meijerink2007} recently examined chemical evolution of Xray-
and UV- irradiated molecular gas with a wide variety of input
parameters.
They calculated thermal and chemical evolutions of plane parallel gas
slabs.
Their results show that in a photodissociation region (PDR) the ratio
tends to be high ($R_{\mathrm{HCN/HCO}^{+}}>1$) but in an Xray
dominated region (XDR) the tendency is reverted
($R_{\mathrm{HCN/HCO}^{+}}<1$), except for a small number of models of
high density and strong Xray emission.
Though the majority of these kinds of calculations assumes a simple
geometry for molecular gas \citep[e.g.,][]{maloney96, meijerink2005,
  meijerink2007}, PDR and XDR chemistry currently does not seem to
explain the ``HCN diagram'' classification \citep{kohno2005} of AGN
and starbust galaxies.
As for alternative scenarios, the intense ionization flux from nearby
supernova remnant shock waves (and accordingly high ionization
degree, \citealt{lepp96}) and shock chemistry have been proposed
\citep[e.g., ][]{nguyen1992}, but it has been still uncertain whether these ISM
chemistry models can find the evolutinonary path for yielding
$y_\mathrm{HCN}/y_{\mathrm{HCO}^{+}} > 1$.
Besides the chemical abundance models, \citet{aalto2007} discussed the
possible role of MIR photons from the dusty torus in determining the
excitation conditions of the emitting molecules for LIRG/ULIRGs.

We expect that the molecular torus should have an inhomogeneous
structure, and then the shielding of high energy (UV and Xray) photons
that affects the chemical and thermal structures would be
significantly different from the those predicted by one-zone or
one-dimensional cloud models.
Three-dimensional chemical abundance effects, such as simple models of
PDR and XDR will be investigated with the three-dimensional scheme in the
subsequent paper.

%--------------
\subsection{Implications for Future Observations}

We mainly examine molecular line emission in millimeter band on the
basis of the nonLTE radiation transfer simulation so far.
If we consider a nearby galaxy at the distance $D = $ 20 Mpc,
the size of the bright spots in our results (Fig.\ref{fig:intmap}), of
which radius $\approx 1$ pc, corresponds to the angular size
$\Delta \theta \approx 0.03^{\prime\prime}$.
Therefore internal structures of a torus will be resolvable 
with ALMA in the line intensity distributions. 
The dispersions of $N_{\mathrm{H}_2}$, $\alpha_\nu$, and the
intensities presented in Figures \ref{fig:tau_o1} and \ref{fig:tau_o}
will be revealed in detail by ALMA's high angular resolution
observation as well. 
We briefly discuss the properties relevant to future observations.

Equation (\ref{eq:lprofile}) means that our radiative transfer
calculations take into account the turbulent velocity field inside
the molecular torus.
Figure \ref{fig:lprofile} demonstrates the line profiles with 10 pc
spacing.
We assume a nearby galaxy at the distance $D\sim 20$ Mpc and take
the binning size of the velocity component along the line of sight to
16 km s$^{-1}$.
The line profiles in Figure \ref{fig:lprofile} show an extremely
complicated structure, reflecting the inhomogeneous structure and
the turbulent velocity field in the torus.
Since a line of sight can pass through more than one dense 
clump, multi peaks appear in the line profile, even if
the molecular abundance is uniform.
Figure \ref{fig:lprofile2} presents the average profile over the
entire field of view.
The compiled profile in Figure \ref{fig:lprofile2} shows a
Gaussian-like structure of single components: this means that even if
current observational results present Gaussian-like profiles, 
future high-resolution observations will be able to reveal the
internal substructures of the compact molecular torus ($R\lesssim 100$
pc).

%------
\section{Summary \& Conclusion}

We performed three-dimensional nonLTE radiative transfer calculations
for HCN and HCO$^{+}$ rotational lines based on high-resolution
hydrodynamic simulation of an AGN molecular torus.
An AGN molecular torus is expected to have exceedingly
inhomogeneous density and temperature structures, and accordingly
complicated energy level populations. 
In this article we examined the nonLTE level population distribution
in the inhomogeneous molecular torus and its effects on the line ratio
$R_{\mathrm{HCN/HCO}^{+}}$ with the assumption of the spatially
uniform chemical abundance distribution.
The results are summarized as follows: 
1) The ratio of HCN and HCO$^{+}$ rotational lines becomes
$R_{\mathrm{HCN/HCO}^{+}}\lesssim 1$ for a wide range of molecular 
abundance $y$ ($10^{-11}\le y_\mathrm{HCN}=y_{\mathrm{HCO}^{+}}
\le 10^{-7}$), thus obtaining a high
ratio $R_{\mathrm{HCN/HCO}^{+}}\sim 2$ observed in some galaxies
(such as NGC 5194 and NGC 1068, \citealt{kohno2005}) requires
more abundant HCN than HCO$^{+}$ ($\langle
y_\mathrm{HCN}\rangle \gtrsim 10\times \langle
y_{\mathrm{HCO}^{+}}\rangle$).
2) The spatial distribution of line ratio $R_{\mathrm{HCN/HCO}^{+}}$
is inhomogeneous, reflecting the spatially inhomogeneous structure of the
molecular torus.
3) Inhomogeneity in the structure of the molecular torus generates a 
dispersion around the linear relations between intensity and
column density.
The stimulated emission by population inversion can dominate the
integrated intensity where $n_\mathrm{{H}_2}\sim n_\mathrm{crit}$
(panel $c$ of Fig.\ref{fig:tau_o}).
When the gas is subthermally populated and marginally thick
($\tau_0\sim 1$), the line ratio may indicate neither the fraction of
high density molecular gas nor the ratio of surface area of $\tau_0=1$
region.
Furthermore, our three-dimensional nonLTE calculations demonstrate the
complex line profile distribution in the synthetic image of the torus
(Fig. \ref{fig:lprofile}),
even with the uniform abundance assumption.
The forthcoming ALMA's high-resolution observations will reveal
such complex structures.
The compilation of high-resolution hydrodynamic simulations and
three-dimensional radiative transfer calculations 
will open a powerful way to study the compact molecular gas in central
regions of external galaxies.

%----
\acknowledgments

We are grateful to K. Kohno and K. Nakanishi for their helpful comments
on the millimeter observations, and to M. Imanishi for his comments on
IR observations of LIRGs and ULIRGs.
MY thanks K. Sakamoto and S. Takakuwa for their advice on the analysis
of radiative transfer results.
This research was supported in part by Grant-in-Aid by the Ministry of
Education, Science, and Culture of Japan (16204012, 17340059, 18026008).

%----
\appendix

\section{Excitation Temperature Distribution}

In the Appendix we describe the fundamental mechanisms that determine
nonLTE level population or excitation temperature distribution in the
optically thin limit.
We especially investigate the superthermal level population appearing
at $n \sim n_\mathrm{crit}$, because the stimulated emission due
to the superthermal population plays an important role for the line
intensities of high density tracers.
In order to provide a simple description of the populating processes,
we here focus on the low energy level excitation in terms of the ratio
$n_1/n_0$, compared with the two level system.

%-----
\subsection{Population Distribution in the Optically Thin Limit}
\label{sect:thinpop}

We first study how the ratio $n_1/n_0$ behaves according to the
increase in the number density of collisional excitation partner
$n_{\mathrm{H}_2}$ in the optically thin limit.
In Figure \ref{fig:f10_den2} the fractional population $n_1/n_0$ is
plotted as a function of density for various temperatures.
We solved equations (\ref{eq:pop}) and (\ref{eq:pop2}) ignoring the
average line intensity $\bar{J}$ except for the cosmic microwave
background radiation ($\bar{J}= B_{\nu}(T_\mathrm{CMB}) \equiv
J_\mathrm{CMB}$).
In Figure \ref{fig:f10_den2},  HCO$^{+}$ and HCN populations are shown
in black and blue lines, respectively.
It is obvious in Figure \ref{fig:f10_den2} that the fractional population
$n_1/n_0$ of HCN coincides with that of HCO$^{+}$ at the extremities of
low ($n_{\mathrm{H}_2}\lesssim 10^2$ cm$^{-3}$) and
high ($n_{\mathrm{H}_2}\gtrsim 10^7$ cm$^{-3}$) densities for the same
temperature.
This fact can be easily understood in terms of the multi-level
population of pure rotational transitions as follows:

In the low density limit, collisional excitation by molecular
hydrogen is negligible, and level population distribution can be
calculated by the balance of radiative excitation by CMB photons 
and spontaneous emission decay (Eq.[\ref{eq:a_thin}]).
Inserting $\bar{J}=J_\mathrm{CMB}$ in equation (\ref{eq:a_thin}),
the ratio $n_1/n_0$ becomes constant for both of HCN and HCO$^{+}$, 
\begin{equation}
  \frac{n_1}{n_0} =
  \frac{g_1/g_0B_{10}J_\mathrm{CMB}}{A_{10}+B_{10}J_\mathrm{CMB}}
   = 0.67.   \label{eq:thinlim10}
\end{equation}
In the second equality in equation (\ref{eq:thinlim10}) we use the
fact that Einstein's $A$ and $B$ coefficients for HCN and HCO$^{+}$
rotational transitions are almost the same (Table \ref{tbl:coef}).

On the other hand, in the high density limit, the level population
converges to the Boltzmann distribution,
\begin{eqnarray*}
  \frac{n_J}{n_{J^{\prime}}} 
    &=& \frac{g_J}{g_{J^{\prime}}}
      \exp\left( -\frac{E_J-E_{J^{\prime}}}{k_BT} \right),  \\
  &=& \frac{2J+1}{2J^{\prime}+1}
      \exp\left[ 
       -\frac{Bh\left\{ J(J+1)-J^{\prime}(J^{\prime}+1)\right\}}{k_BT} 
      \right].  
\end{eqnarray*}
From Table \ref{tbl:coef}, rotational constants $B$(HCN) $\simeq$
$B$(HCO$^{+}$), and then the Boltzmann distributions of two lines are
in perfect agreement, 
\begin{equation}
  \left(\frac{n_1}{n_0}\right)_\mathrm{LTE} 
   = 3\exp\left( -\frac{2Bh}{k_BT}\right). \label{eq:10lim}
\end{equation}
Equations (\ref{eq:thinlim10}) and (\ref{eq:10lim}) explain the
natural concordance of $n_1/n_0$ of HCN$(1-0)$ and HCO$^{+}(1-0)$ in
both low and high density extremities.

Figure \ref{fig:f10_den2} exhibits a peak of $(n_1/n_0)$ higher than
that expected for LTE $(n_1/n_0)_\mathrm{LTE}$ at the intermediate
density ($10^2$ cm$^{-3} \lesssim n_{\mathrm{H}_2} \lesssim 10^7$
cm$^{-3}$) for each temperature.
This peak is a result of a combination of two effects, namely the
energy level cascade in a multi-level system, and the strong $J$
dependence of critical densities for pure rotational transitions.

The energy level cascade originates in the selection rule of radiative
transitions.
Pure rotational transitions have a selection rule of $\Delta J = \pm
1$. 
The selection rule significantly simplifies the rate equation in the
low density and optically thin limit \citep{yamada2007},
  \begin{equation}
  n_JA_{J, J-1} = \sum_{J^{\prime}\ge J}C_{0,J^{\prime}}n_0. 
     \label{eq:rate_thin}
  \end{equation}
Equation (\ref{eq:rate_thin}) means that a collisionally excited
particle from the ground state should cascade down all the steps
of the energy level ladder with $\Delta J=-1$ accompanied by
spontaneous emission photons.

As density increases, collisional de-excitation begins to modify 
the simple form of rate equation (\ref{eq:rate_thin}).
As shown in equations (\ref{eq:einsB}) and (\ref{eq:einsA}), the
Einstein's $A$ coefficient for pure rotational transition is strongly
dependent on $J$ as $A_{J, J-1}\propto J^3$.
On the other hand, the collisional transition constant only weakly 
depends on $J$ \citep{goldreich74, mckee82}.
Thus the critical density for LTE $n_\mathrm{crit}(J, J-1) \equiv
A_{J, J-1}/\gamma_{J,J-1}$ sharply increases with $J$ ($\propto J^3$).
Figure \ref{fig:hcn_test} displays the fractional level population
distributions $f_J$ ($f_J\equiv n_J/n_\mathrm{mol}$, $\sum_J f_J = 1$)
as a function of $J$ for various densities ($n_{\mathrm{H}_2}=10^5$,
$10^7$, and $10^9$ cm$^{-3}$).
The overall distribution of $f_J$ approaches the Boltzmann
distribution (indicated with a solid line in Fig. \ref{fig:hcn_test})
as the density increases.
In Figure \ref{fig:hcn_test}, it is also shown that while $f_J$ is close
to the Boltzmann distribution for low $J$, $f_J$ deviates from the
Boltzmann distribution for high $J$.
This is another expression for the reason the peak of $n_1/n_0$
appears (Fig. \ref{fig:f10_den2}): for a fixed density $n$, the low
$J$ levels that satisfy $n\gg n_\mathrm{crit}(J, J-1)$ approaches the
Boltzmann distribution (the solid line in Fig. \ref{fig:hcn_test}),
and the further downward level cascade from the corresponding level
comes to a halt.
However, at higher $J$ levels where the LTE condition is not
satisfied, the downward cascade by spontaneous emission 
accumulates the cascading particles at the density
$n\sim n_\mathrm{crit}$.
Accumulation of the extra cascaded particles in addition to the
Boltzmann distribution generates a peak of $n_1/n_0$ shown in
Figure \ref{fig:f10_den2}.

%......
\subsection{Population Inversion} \label{sect:popinv}

The source function $S_\nu$ and the absorption coefficient
$\alpha_\nu$ from the level $J$ to $J-1$ denote
\begin{eqnarray}
  S_\nu &=& \frac{2h \nu^3}{c^2}
            \left( \frac{g_J}{g_{J-1}} \frac{n_{J-1}}{n_J} -1
                   \right)^{-1},  \\
  \alpha_\nu &=& \frac{h\nu}{4\pi} n_{J-1}B_{J-1, J}\phi(\nu)
            \left(1-\frac{g_{J-1}}{g_J} \frac{n_J}{n_{J-1}} \right).
            \label{eq:alpha}
\end{eqnarray}
These equations imply negative $S_\nu$ and $\alpha_\nu$ if
$\delta_\mathrm{pop}\equiv n_Jg_{J-1}/n_{J-1}g_J$ exceeds unity
(population inversion: \citealt{rybicki1979}).
In Figure \ref{fig:popinv} we plot $\delta_\mathrm{pop}$ for
HCN$(1-0)$ and HCO$^{+}(1-0)$ in the optically thin limit as a
function of density and temperature.
In our torus model, density and temperature range $n_{\mathrm{H}_2}
\lesssim 2\times 10^6$ cm$^{-3}$ and 20 K $\le T_\mathrm{kin} \le $ 1000
K, respectively.
Figure \ref{fig:popinv} suggests $\delta_\mathrm{pop}$ is likely to
become larger than unity in this density and temperature regime.

When population inversion occurs in the $i-$th grid
($\delta_\mathrm{pop}^i > 1$), optical thickness in the grid $\Delta
\tau_\nu^i = \alpha_\nu^i \Delta s$ becomes negative due to negative
absorption coefficient $\alpha_\nu^i$.
It is apparent in terms of the formal solution of radiative transfer
$I_\nu(\tau_\nu) = I_\nu(0)e^{-\tau_\nu}+S_\nu(1-e^{-\tau_\nu})$ that
negative optical thickness $\tau_\nu < 0$ strongly enhances the
specific intensity $I_\nu$.
Since population inversion can significantly affect the intensity, it
should be treated correctly in line transfer calculations.
As described in \S\ref{sect:thinpop}, one of the origins of population
inversion between $J$ and $J-1$ (or peak of $n_1/n_0$ in
Fig.\ref{fig:f10_den2}) is the downward energy level cascade from
higher levels $j$ $(>J)$ to $J$.
Therefore the maximum energy level $J_\mathrm{max}$ in the radiative
transfer calculation has to be as large as possible to obtain the
precise level population, especially when population inversion is
expected.
Preliminary radiative transfer calculations 
reveal the lower limit of the maximum energy level should be 
$J_\mathrm{max}\ge$ 8 in our model torus.
In this paper, we take the upper limit of $J_\mathrm{max} = $10.

Intense population inversion ($\delta_\mathrm{pop}\ge 1$) occasionally
takes place during the iterative calculations before the solution of
level population meets the final convergence.
This sort of temporal population inversion sometimes induces numerical
divergence of the intensity due to a factor $e^{-\tau_\nu}$.
We avoid the temporal numerical divergence using a way similar to 
that of \citet{hogerheijde2000}: we set the numerical upper bounds on
$\bar{J}$ and $|\tau_\nu|$.
Numerical experiments with several sets of upper bounds present an
excellent convergence, thus confirming the stability of the obtained
solutions against the variation of the numerical limiters.

%----

\clearpage

%% insert figures.

% -- fig1
%
% -- fig2
%
% -- fig3
%
% -- fig4
\begin{center}
  \begin{figure}
    \includegraphics[width=7cm,clip]{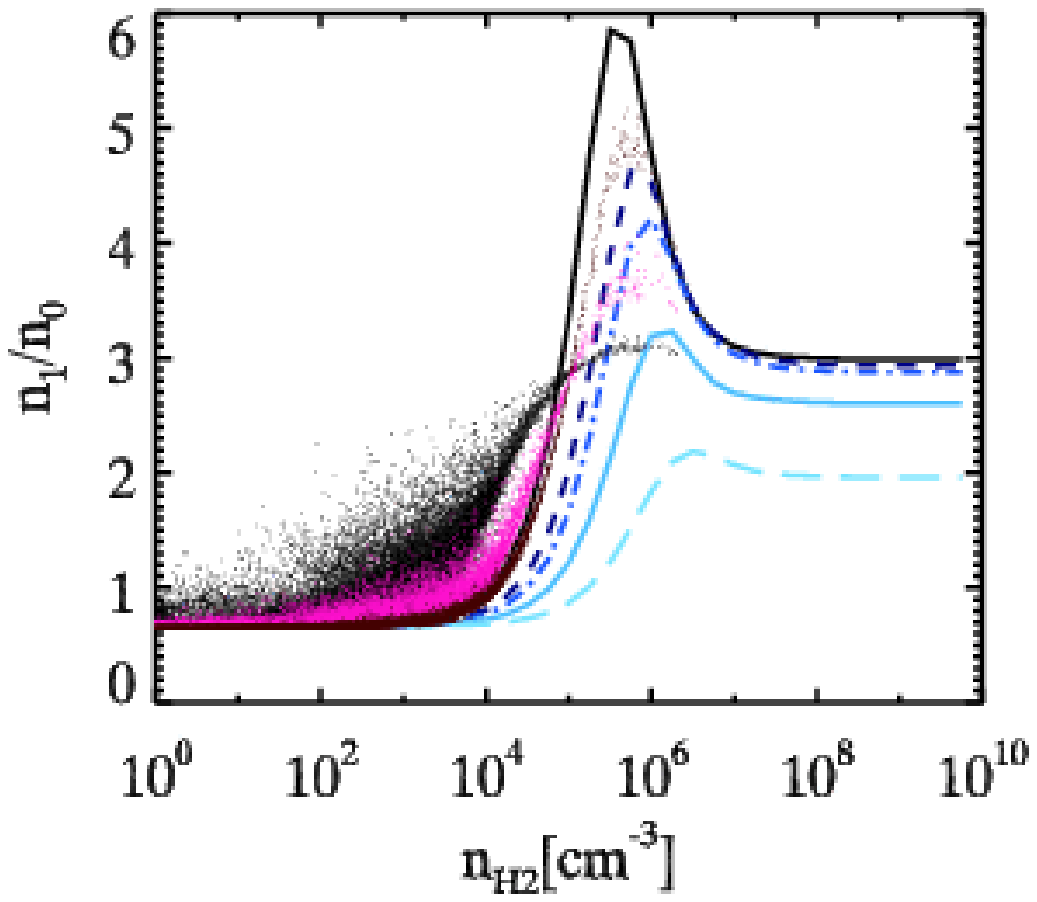}
    \includegraphics[width=7cm,clip]{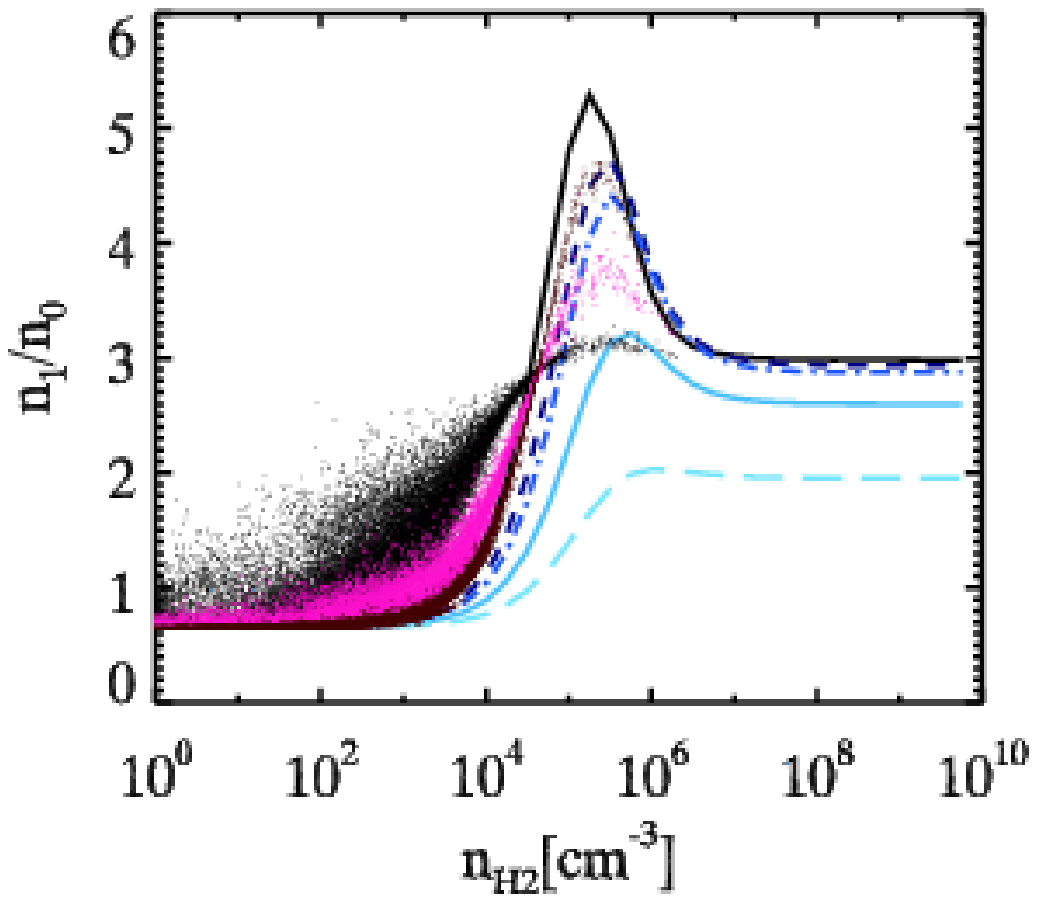}
    \caption{The relative population $n_1/n_0$ as a function of
      density $n_{\mathrm{H}_2}$ for HCN (the left panel) and
      HCO$^{+}$ (the right panel).
      Lines are those for the optically thin limit (the same as Figure
      \ref{fig:popinv}) of which kinetic temperatures are 10, 30, 100,
      200, and 1000 K, respectively.
      The results of our radiative transfer calculations for three
      values of $y$ are presented with dots of different colors:
      black dots are of $y=2\times 10^{-8}$, pink dots are of
      $y=2\times 10^{-9}$, and dark red dots are of $y=2\times
      10^{-10}$. 
        }
    \label{fig:fj10_sim}
  \end{figure}
\end{center}
%
% -- fig5
\begin{center}
  \begin{figure}
    \includegraphics[width=7cm]{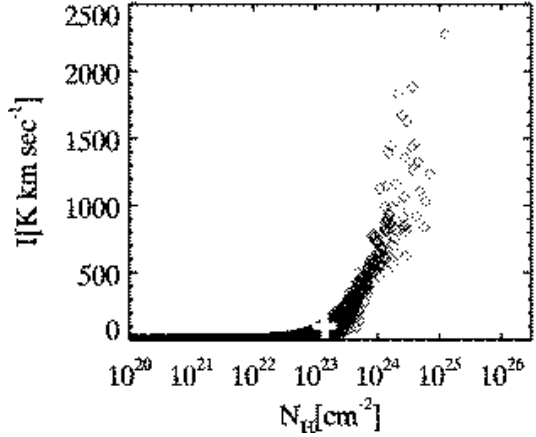}
    \caption{ The integrated intensity as a function of
    the column density $N_{\mathrm{H}_2}$ of the face-on data. 
    When $N_{\mathrm{H}_2}$ is larger than $\sim 10^{23}$
    cm$^{-2}$, the dispersion of the integrated intensity $I$
    becomes accordingly larger.
    A cross at $N_{\mathrm{H}_2}=9.27\times 10^{22}$ cm$^{-2}$ and $I =
    42.0$ K km s$^{-1}$ indicates the average column density and the
    average integrated intensity in the field of view. }
    \label{fig:tau0}
  \end{figure}
\end{center}
%
% -- fig6
\begin{center}
  \begin{figure}
    \includegraphics[width=7.2cm, clip]{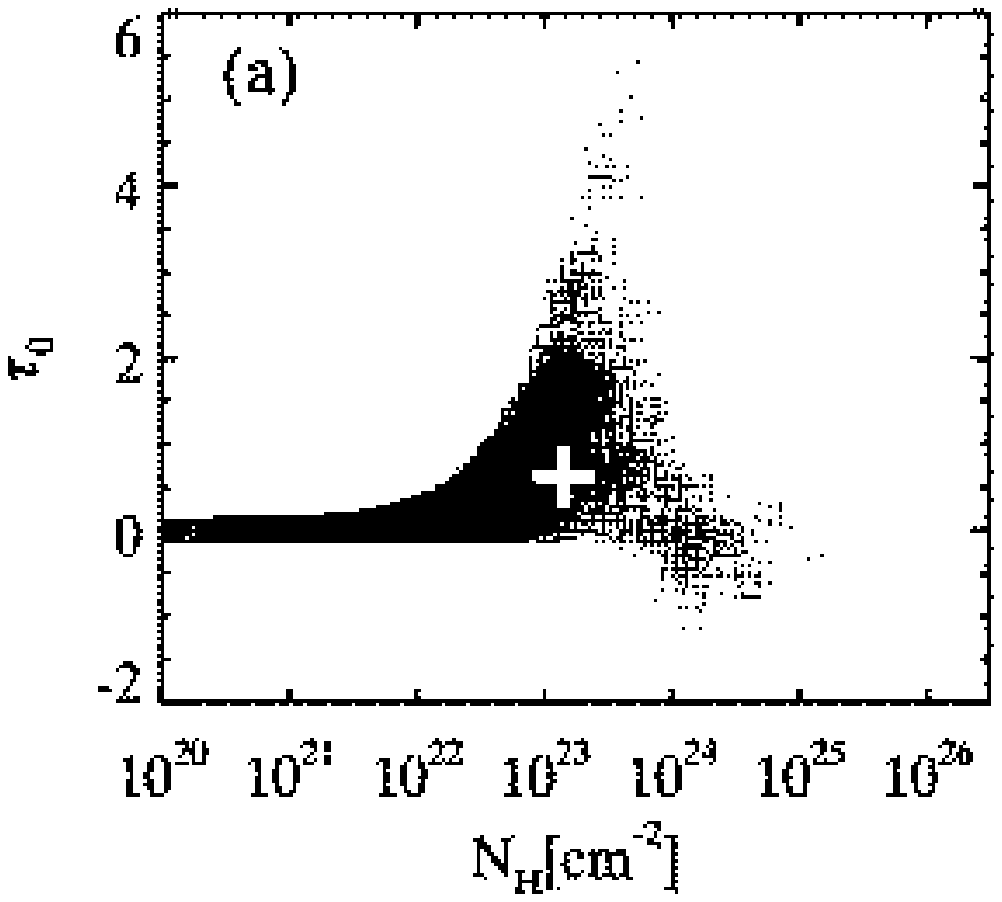}
    \includegraphics[width=8cm, clip]{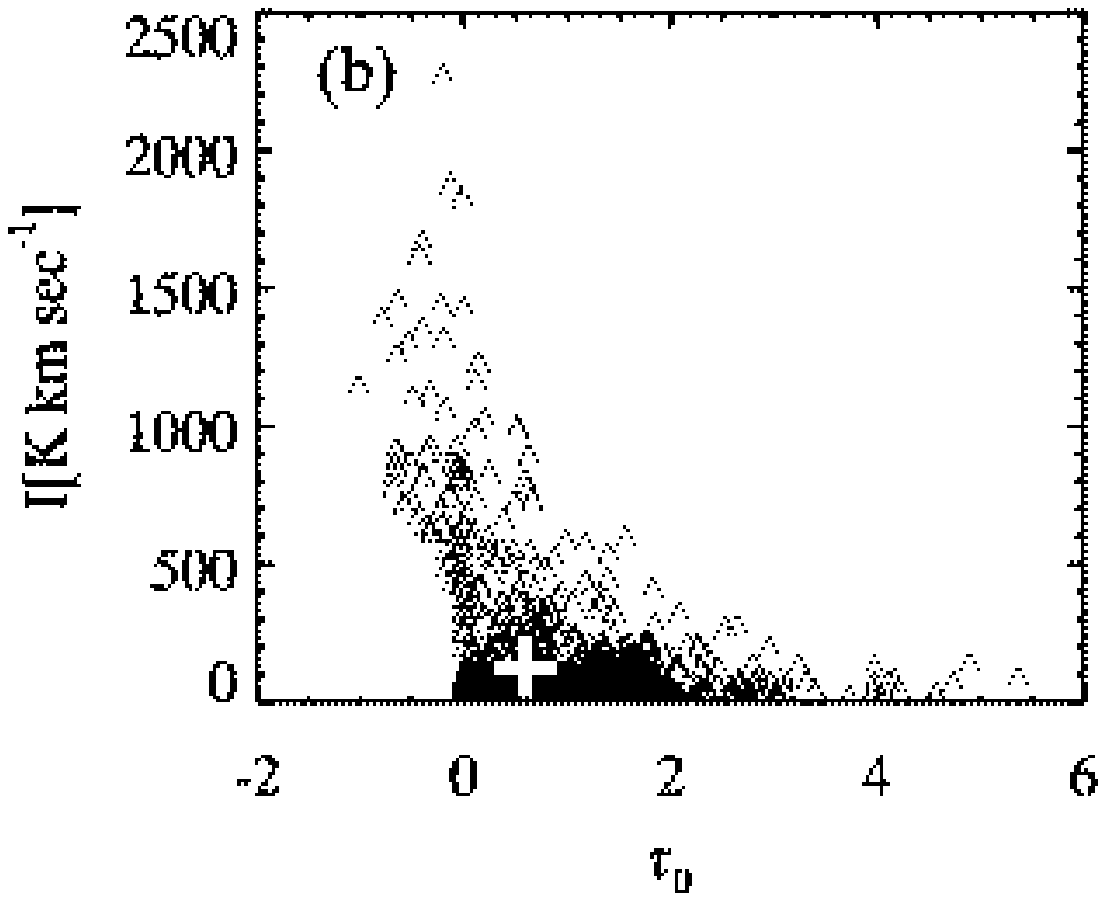}
    \caption{The distribution of optical thickness $\tau_0$ and 
      column density $N_{\mathrm{H}_2}$ (panel $a$), and of 
      optical thickness and integrated intensity (panel $b$).
      The data are taken from the results of HCN$(1-0)$ line calculation
      with $y = 2\times 10^{-9}$.
      Crosses at $(N_{\mathrm{H}_2}, \tau_0) = (9.27\times 10^{22}
      ~[\mathrm{cm}^{-2}], 0.37)$ in panel $(a)$ and $(\tau_0, I) =
      (0.37, 42.0$ [K km s$^{-1}$]) in panel $(b)$ are the averaged
      values in the field of view, respectively.
      The dispersion of the optical thickness $\tau_0 = \int
      \alpha_{\nu_0} ds$ and the lack of proportionality to the column
      density arise from the large dispersion of the absorption
      coefficient $\alpha_\nu$, which is calculated by the nonLTE
      level population (eq. [\ref{eq:alpha}]) in the inhomogeneous
      molecular torus (see text).
    }
    \label{fig:tau_nh}
  \end{figure}
\end{center}
%
% -- fig7
\begin{center}
  \begin{figure}
    \includegraphics[width=8cm, clip]{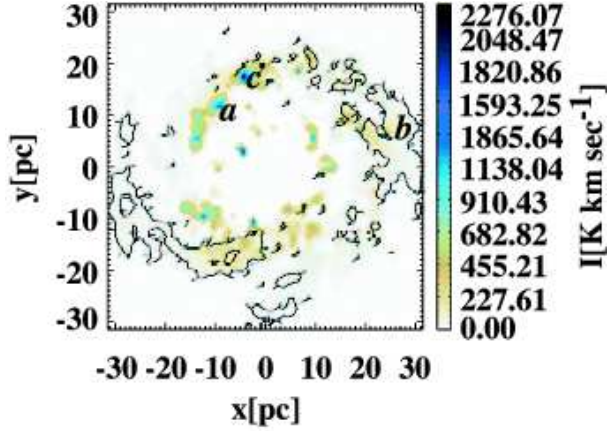}
    \caption{Distribution of the integrated intensity of HCN$(1-0)$ line
    overplotted by the contour of $\tau_0 \ge 1$.
    The data are of $y = 2\times 10^{-9}$.
    One can observe discordance of the bright regions (color scale)
    and the optically thick regions (see text).
    }
    \label{fig:tau_o1}
  \end{figure}
\end{center}
%
% -- fig8
\begin{center}
  \begin{figure}
    \includegraphics[width=17cm, clip]{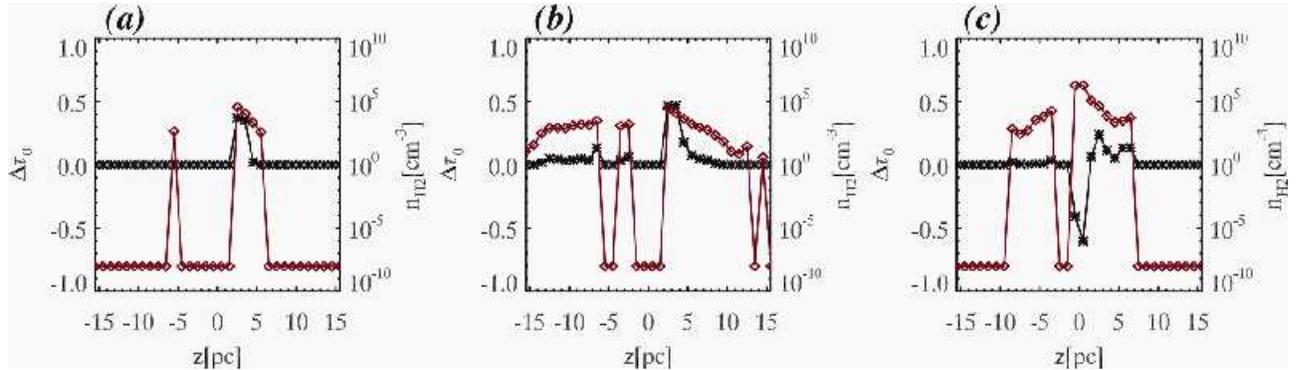}
    \caption{
    The distributions of $\Delta\tau_0$
    (lines with crosses) and the density (lines with diamonds) along
    three lines of sight indicated in Fig. \ref{fig:tau_o1}. 
    The lowest density ($n_{\mathrm{H}_2}= 10^{-8}$ cm$^{-3}$) seen in
    these panels is the cut-off density adopted in the hydrodynamic
    simulation.
    These three panels represent different circumstances: in panel
    $(a)$ emission from a dense clump at $z\simeq 3$ is seen
    through a tenuous ambient, in panel $(b)$ the optical thickness
    is large due to accumulation of
    $\Delta \tau_0$ of low density atmosphere encompassing the large
    scale height, and in panel $(c)$ the intensity is strong because
    of the stimulated emission by the population inversion within a
    dense region ($n_{\mathrm{H}_2}\lesssim n_\mathrm{crit}$) around
    $z=0$, respectively.
    }
    \label{fig:tau_o}
  \end{figure}
\end{center}
%
% -- fig9
\begin{center}
  \begin{figure}
    \includegraphics[width=14cm]{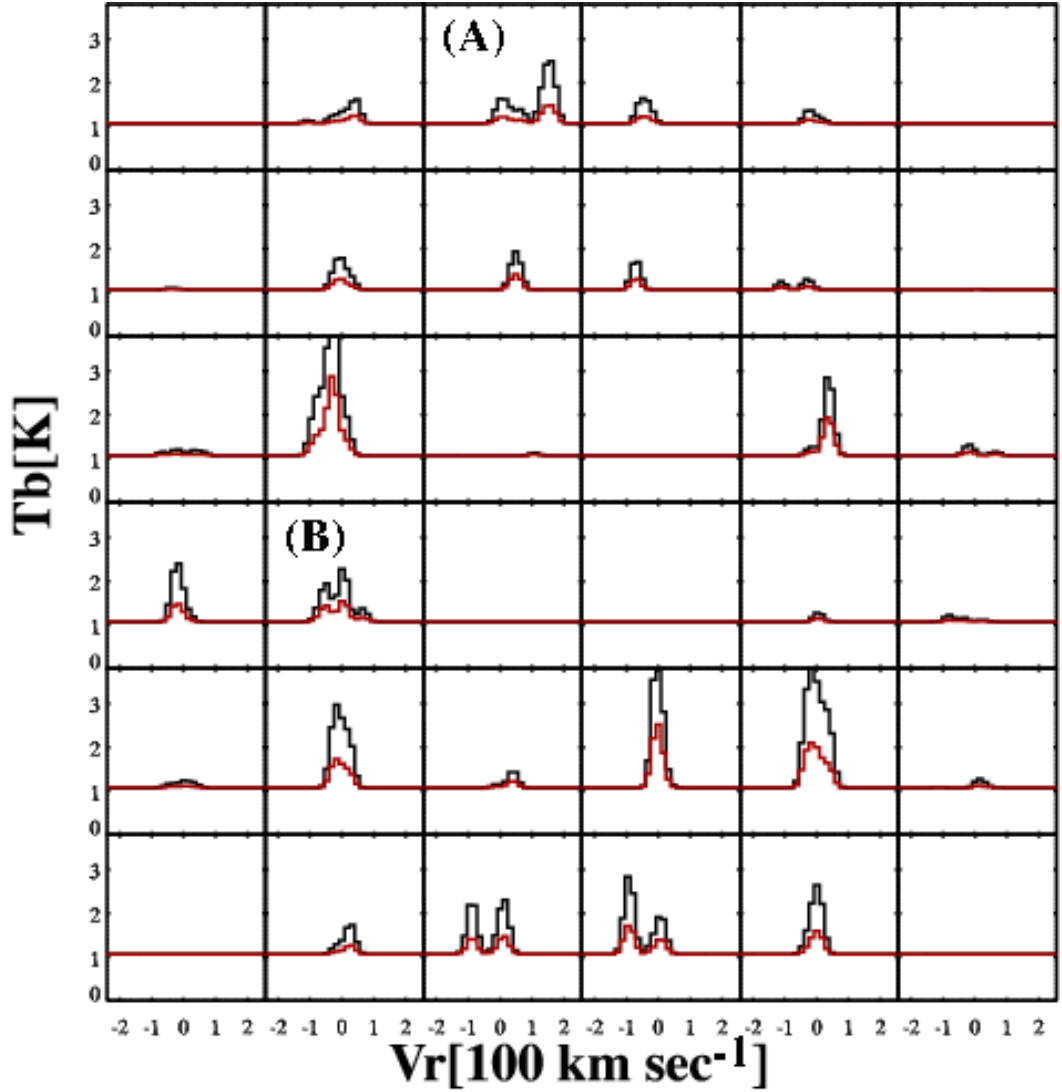}
    \caption{The distribution of line profiles at the points with 10
    pc spacing.
    Abscissa denotes the velocity component along the line of sight
    measured from the position of the torus.
    The binning size of $V_r$ is taken to be 16 km s$^{-1}$.
    In each panel, black lines are HCO$^{+}(1-0)$ and red lines
    are HCN$(1-0)$ lines.
    Multi-peak profiles at the points labeled $(A)$ and $(B)$ reflect
    the inhomogeneous structure of the torus.}
    \label{fig:lprofile}
  \end{figure}
\end{center}
%
% -- fig10
\begin{center}
  \begin{figure}
    \includegraphics[width=6cm]{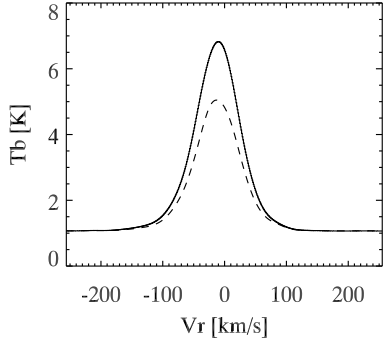}
    \caption{The averaged line profiles over the whole field of view.
    The solid line denotes HCO$^{+}$, and the dashed line denotes HCN
    profile.
    The inhomogeneous distribution of line profiles seen in
    Fig. \ref{fig:lprofile} is compiled to a
    single Gaussian-like profile with width $\Delta v \sim
    200$ km s$^{-1}$.}
    \label{fig:lprofile2}
  \end{figure}
\end{center}
%
% -- fig11
\begin{center}
  \begin{figure}
    \includegraphics[width=7cm]{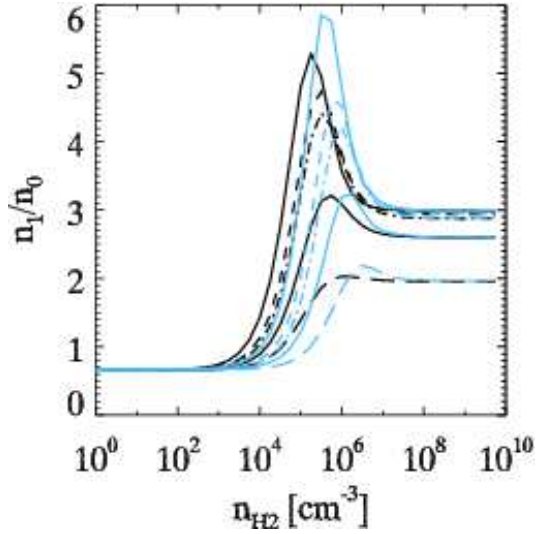}
    \caption{Density dependence of the ratio $n_1/n_0$ (the excitation
    temperature $T_\mathrm{ex}$).
    Blue lines represent HCN, and black lines do HCO$^{+}$ transitions
    for various kinetic temperatures: long-dashed
    lines are $T$ = 10K, three-dot-dashed lines are $T$ = 30K,
    dot-dashed lines are $T$ = 100K, short-dashed lines are $T$ = 200K,
    and solid lines are $T$ = 1000K.
    In both of the extremities of low and high density, $n_1/n_0$ for
    HCN and HCO$^{+}$ molecules coincide.
    See Appendix for details.}
    \label{fig:f10_den2}
  \end{figure}
\end{center}
%
% -- fig12
\begin{center}
  \begin{figure}
    \includegraphics[width=7cm,clip]{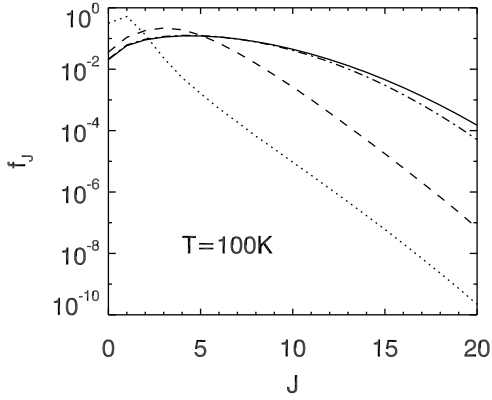}
    \caption{The fractional level population ($f_J$) distributions of
      HCN rotational line for various densities in the optically thin
      limit with $T=100$ K.
      In this panel, the dotted lines denote $n_{\mathrm{H}_2}=10^5$
      cm$^{-3}$, the dashed lines $n_{\mathrm{H}_2}=10^7$
      cm$^{-3}$, the dot-dashed lines $n_{\mathrm{H}_2} = 10^9$
      cm$^{-3}$, and the solid lines the Boltzmann distribution.
        }
    \label{fig:hcn_test}
  \end{figure}
\end{center}
%
% -- fig13
\begin{center}
  \begin{figure}
  \includegraphics[width=10cm, clip]{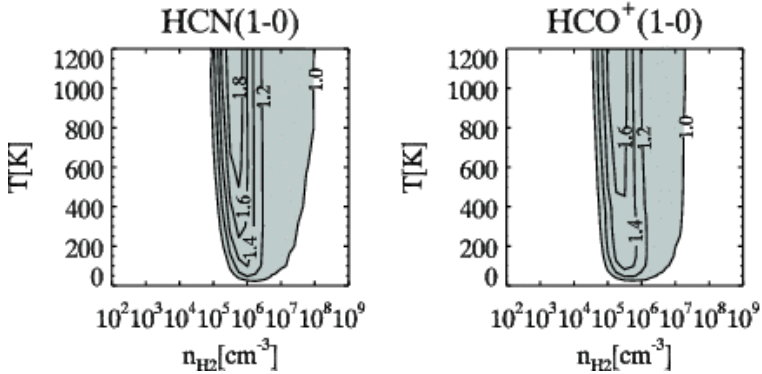}
  \caption{The degree of population inversion $\delta_\mathrm{pop}$ as
  a function of temperature and density for HCN$(1-0)$ (left panel)
  and HCO$^{+}(1-0)$ (right panel) transitions which is calculated in
  the optically thin limit.
  Contours are from $\delta_\mathrm{pop} =$  1.0 to 1.8 with 0.2
  spacing, and population inversion occurs within the outermost
  contour (gray regions).}
  \label{fig:popinv}
  \end{figure}
\end{center}

\end{document}